\documentclass[manuscript,screen, acmsmall]{acmart}
\usepackage{booktabs}
\usepackage{caption}
\usepackage{subfigure} 
\usepackage{tabularx}
\usepackage{color, colortbl}
\usepackage{xcolor}
\usepackage{siunitx}
\usepackage{multirow}

\newcolumntype{L}[1]{>{\raggedright\let\newline\\\arraybackslash\hspace{0pt}}m{#1}}
\newcolumntype{C}{>{\centering\arraybackslash}S}
\newcolumntype{R}[1]{>{\raggedleft\let\newline\\\arraybackslash\hspace{0pt}}m{#1}}

\AtBeginDocument{%
  \providecommand\BibTeX{{%
    \normalfont B\kern-0.5em{\scshape i\kern-0.25em b}\kern-0.8em\TeX}}}

\copyrightyear{2024}
\acmYear{2024}
\setcopyright{acmcopyright}\acmConference[CHI '24]{Proceedings of the CHI Conference on Human Factors in Computing Systems}{May 11--16, 2024}{Honolulu, HI, USA}
\acmBooktitle{Proceedings of the CHI Conference on Human Factors in Computing Systems (CHI '24), May 11--16, 2024, Honolulu, HI, USA}
\acmDOI{10.1145/3613904.3642269}
\acmISBN{979-8-4007-0330-0/24/05}

\begin{document}

\title[Founder Motivations, Goals, and Actions]{How Founder Motivations, Goals, and Actions Influence Early Trajectories of Online Communities}

\author{Sanjay R. Kairam}
\email{sanjay.kairam@gmail.com}
\orcid{0000-0001-8320-222X}
\affiliation{%
  \institution{Reddit, Inc.}
  \city{San Francisco}
  \state{California}
  \country{USA}
  \postcode{94104}
}

\author{Jeremy Foote}
\email{jdfoote@purdue.edu}
\orcid{0000-0002-0078-2925}
\affiliation{%
  \institution{Brian Lamb School of Communication, Purdue University}
  \city{West Lafayette}
  \state{Indiana}
  \country{USA}
  \postcode{47907-2050}
}

\renewcommand{\shortauthors}{Sanjay R. Kairam and Jeremy Foote}

\begin{abstract}
    Online communities offer their members various benefits, such as information access, social and emotional support, and entertainment. Despite the important role that founders play in shaping communities, prior research has focused primarily on what drives \textit{users} to participate and contribute; the motivations and goals of \textit{founders} remain underexplored. To uncover how and why online communities get started, we present findings from a survey of 951 recent founders of Reddit communities. We find that topical interest is the most common motivation for community creation, followed by motivations to exchange information, connect with others, and self-promote. Founders have heterogeneous goals for their nascent communities, but they tend to privilege community quality and engagement over sheer growth. These differences in founders' early attitudes towards their communities help predict not only the community-building actions that they pursue, but also the ability of their communities to attract visitors, contributors, and subscribers over the first 28 days. We end with a discussion of the implications for researchers, designers, and founders of online communities.
\end{abstract}

\ccsdesc[500]{Human-centered computing~Empirical studies in collaborative and social computing}
\ccsdesc[500]{Information systems~Collaborative and social computing systems and tools}

\keywords{Reddit, online communities, motivations, community creation, founders}

\received{14 September 2023}
\received[revised]{14 November 2023}
\received[accepted]{19 January 2024}

\maketitle

\section{Introduction}
Online communities provide multiple benefits for their members, including opportunities to share and receive information~\cite{iriberri_life-cycle_2009, lampe2010motivations, moore2017redditors, ridings2006psychological, wang2003assessing, matthews_goals_2014}, to connect and socialize with others~\cite{iriberri_life-cycle_2009, lampe2010motivations, moore2017redditors, wang2003assessing, matthews_goals_2014}, to be entertained~\cite{lampe2010motivations, moore2017redditors}, to exchange social or emotional support~\cite{iriberri_life-cycle_2009, ridings2006psychological, saha2020understanding}, and to gain status or prestige~\cite{moore2017redditors, wang2003assessing, matthews_goals_2014}. On massive community-based platforms such as Reddit, communities cover a broad range of topics, serve multiple audiences~\cite{waller2019generalists, waller2021quantifying}, and manage themselves using diverse community-determined rules and norms~\cite{fiesler2018reddit, chandrasekharan2018internet}. Before communities can provide any of these benefits to their members, however, they must first achieve `critical mass'~\cite{granovetter1978threshold, oliver1985theory, grudin1994groupware, raban2010empirical}, or a large enough set of initial members to make membership worthwhile for newcomers.

Founders of new communities play a central role in building these spaces: they set the topic of the community, often set specific rules and guidelines, and recruit others to join. Their actions within the early days of a community can make the difference between whether a new community will thrive or fail~\cite{kraut2014role}. Which actions they choose to take are influenced by the affordances of the platform, their own motivations for founding a community, and their goals for their community. While much of the literature targeted at researchers and practitioners implicitly frames community success as growing `larger' (e.g.~\cite{backstrom2006group, butler2001membership, cunha2019all, iriberri_life-cycle_2009, kairam2012life}, many founders of communities may not prioritize growth~\cite{foote2017starting}, and even small communities can bring specific types of value to their members~\cite{hwang_why_2021, teblunthuis_no_2022}. Despite the prevalence and importance of online communities and the critical role played by founders' motivations and goals in shaping these spaces, very little research has explicitly studied either community founders or early-stage communities. 

In this paper, we study the motivations, goals, and initial strategies of community founders in the context of Reddit, through a survey of 951 recent founders. We find that founder motivations are similar to contributor motivations, with topical interest most prevalent among founders. While founders vary in their goals, a large community is often neither a goal nor an expectation. We find that differences in founder motivations predict differences in their goals and plans to engage in various types of community-building actions. Finally, we analyze how differences in founders' motivations, goals, and planned actions relate to the early trajectories of their communities over the first 28 days. We discuss the implications of these findings not only for those researching online communities and designing tools to support these communities, but also for founders themselves.

\section{Background \& Related Work}
\subsection{Community Motivations}
Prior research has identified several common motivations, which vary by context, for individual participation and contribution in online communities, including information-seeking, social and emotional support, social interaction, entertainment, learning, and status-seeking \cite{lampe2010motivations,iriberri_life-cycle_2009,moore2017redditors}. Researchers have also identified relationships between expressed motivations for participation in online communities and the degree to which individuals participate or contribute. For instance, bulletin board contributors have different motivations than lurkers, with contributors more motivated by the desire to provide information and exchange social support~\cite{ridings2006psychological}. In the Everything2.com online forum, a motivation to provide information was the strongest predictor of self-reported intention to contribute in the future~\cite{lampe2010motivations}. In the context of a travel community, rates of contribution were correlated with motivations to interact socially, express oneself, and provide a service to others, but not with a desire to gain status or ensure that the forum provided high-quality information~\cite{wang2003assessing}. On Reddit, social/community and `status-seeking' motivations, but not information-exchange or entertainment motivations, have been associated with contribution rates~\cite{moore2017redditors}.

User motivations are diverse and crucial in predicting engagement levels; founders undeniably shape the early trajectory and success of their communities~\cite{kraut2014role, resnick_starting_2012}, yet we know almost nothing about the motivations that lead founders to start new communities. While leaders of already-existing communities often report being motivated by `altruism' or `supporting the community'~\cite{saha2020understanding, seering2023moderates}, these motivations are unlikely to be present when a community is first created. The most analogous study, by Foote et al.~\cite{foote2017starting}, found that wiki founders reported motivations like providing useful information, disagreements with existing communities, learning, and promoting personal material. However, the nature of wikis, centered around collaborative content creation, represents only one special case of a broader class of online communities. Our study seeks to bridge this knowledge gap, offering insights into the motivations behind founding diverse, discussion-based communities.

\subsection{Community Goals and Values}

If motivations refer to an individual's reasons for founding or participating in a community, community goals and values capture beliefs about what a founder would like a community to become. As with motivations, most research on goals or values focuses on the members or leaders of existing communities. This research shows that community members and moderators have multiple goals, though common aspects of a community valued by users include community quality, information exchange, and opportunities to connect with others \cite{weld_what_2021, matthews_goals_2014}. Some scholars have thus suggested changes in how we conceptualize community success, such as using multiple metrics or aligning metrics with community goals~\cite{aumayr2017path, preece2001sociability}. Prior studies have focused on users and moderators of existing communities; there are reasons to believe that attitudes to a community may differ at its inception. Prior work has demonstrated that community leaders' goals and their strategies for achieving them shift over time through interactions with community members~\cite{cullen2022practicing}. In the present work, we ask founders about their goals soon after creating a community, before their expectations have been shaped by the experience and work of managing the community.

\subsection{Early Community Trajectories and Founder Influence}
Like any form of collective action, a primary challenge for online communities is attracting a `critical mass' of members, a threshold at which other members' presence creates an incentive for participation~\cite{oliver1985theory, granovetter1978threshold}. Achieving critical mass may depend on attracting contributions at the outset~\cite{raban2010empirical}. This in turn requires motivating potential participants to invest effort into their contributions, often before they are able to realize the full benefits of the community~\cite{grudin1994groupware}. In these early days of a community, there are clear reasons to believe that the founder may play an outsized role in determining whether the community will succeed or fail. 

A broad survey of prior research shows that specific founder strategies, such as raising awareness, welcoming newcomers, encouraging contributions, and regulating behavior can impact the success and survival of a community~\cite{resnick_starting_2012}. Empirical study has shown that founder attributes and actions predict a community's survival. For example, Facebook Groups were more likely to survive when founders had more experience on the platform, were not the only administrators, actively visited and created content, and recruited non-friends to the group \cite{kraut2014role}. In this work, we address the previously unexplored link between founders' attitudes towards their communities and their propensity to take actions that could benefit their community.

\subsection{Research Approach}
To better understand the formation stage of new online communities, we use a survey to measure three aspects of how founders approach their new communities---motivations to create the community, goals for the community, and plans to build the community---and how these influence each other. We then use digital trace data to assess relationships between these attitudes and the number of visitors, contributors, and subscribers to their nascent communities. Specifically, we ask:
\begin{itemize}
    \item \textbf{RQ1} What are the motivations, goals, and community-building plans of online community founders?
    \item \textbf{RQ2} How do motivations, goals, and plans relate to early measures of community activity?
\end{itemize}

To answer RQ1, we use descriptive data and exploratory factor analysis to characterize founder motivations, goals, and plans. For RQ2, we employ regression analyses relating founder attitudes to community growth over the first 28 days. We provide further context via additional analyses showing how motivations, goals, and plans interrelate.

\textbf{Research Context: Reddit.} With over 100,000 active communities and 57 million daily active users around the world, as of 2023~\cite{reddit2023api}, Reddit is an ideal research context for exploring the motivations, goals, and strategies associated with community creation. Each community, or \textit{subreddit}, is created independently and led by Reddit users, known as \textit{moderators}; within the scope of Reddit's overall Content Policy~\cite{reddit2023content} and Moderator Code of Conduct~\cite{reddit2023moderator}, subreddits vary broadly in their topics, rules, and norms~\cite{chandrasekharan2018internet, fiesler2018reddit}. The breadth of communities that we sample from should help to generate findings that will generalize to other types of online communities with similar affordances.

\section{Methods}
A survey was distributed via Qualtrics to a sample of new subreddit creators between March 22-31, 2023, 7 days after they created their new community. Capturing founders' sentiments so soon after the creation of their communities ensures that we are capturing their plans and expectations about the community \textit{before} they have been shaped by their experience moderating these spaces. We contacted 29,065 unique users with respect to 30,838 new subreddits. After removing those excluded due to age restrictions, non-completes, and multiple responses from a small number of users who completed the survey more than once, a total of 951 valid completes remained (3.3\% completion rate). Respondents were compensated \$5 via an Amazon gift card; the median completion time was just under 5 minutes (523 seconds), making the median compensation rate around \$35/hour.

\subsection{Survey Instrument}
The survey instrument (see Appendix) included several sections capturing: (1) informed consent, (2) creator demographics, (3) creator experience, and (4) community-specific measures. In terms of creator experience, participants were asked to self-report tenure and frequency of use for Reddit and their experience participating in, moderating, or starting communities on Reddit or other services. The community-specific section started with questions about the intended topic and audience scope for the new community. To capture motivations, respondents indicated their level of agreement with each of 14 Likert-scale items capturing potential motivations for community creation, drawn from prior studies of motivation for community participation and creation~\cite{foote2017starting, moore2017redditors, park2009being}. 

To capture goal prioritization, respondents were prompted to, ``Think about how you will assess whether or not your community is successful'' and rank a set of seven measures of success drawn from prior work~\cite{foote2017starting}, based on relative importance for their community. Participants had the option to write their own success measure and rank it alongside the others. Finally, participants responded to measures capturing their activities towards launching their community, including how much time they planned to spend weekly on their community. They also indicated their agreement with statements about their plans to pursue several categories of community-building actions, drawn from previous work~\cite{foote2017starting, kraut2012building}---raising awareness, welcoming newcomers, encouraging contributions, and regulating behavior.

\subsection{Survey Participants}
Participants self-reported their age using predefined ranges; the median age was in the `25-34' range. Notably, 29\% (281) indicated they were 35 or older. 89\% of respondents opted to self-disclose their gender; of these, the distribution was 73\% Male, 19\% Female, and 8\% Non-Binary/Other. These self-reported demographics skew younger and more male than the broader Reddit user base~\cite{reddit2022audience}. 

In terms of their tenure on Reddit, 60\% (581) reported that they had used the platform for at least a year. 17\% indicated that they had joined within the last 1-4 weeks, and 4\% indicated tenure of a week or less. 57\% of respondents self-reported that they were first-time moderators on Reddit, 38\% reported having previously created a subreddit, and 6\% reported having previously moderated, but never having created, a subreddit. 50\% reported that they had previously created or moderated a community on another service, outside of Reddit.

Respondents assessed the audience for which their subreddit would be appropriate, based on the maturity level of content shared. 590 (62\%) indicated that their subreddit would be appropriate for Everyone (E), meaning that it ``occasionally or never posts and discusses mature themes.'' The remaining 38\% of subreddits might be considered Mature (M), indicating that the expected content might be unsuitable for a general audience, for any reason (e.g. discussion of certain medical conditions, adult content). 

Respondents reported community topics using a predefined list; Frequently-reported topics were \texttt{Entertainment} (30\%), \texttt{Mature Themes \& Adult Content} (21\%), \texttt{Humor} (20\%), \texttt{Gaming} (18\%) and \texttt{Tech} (15\%). Less common topics included \texttt{Politics/News} (9\%), \texttt{Music} (8\%), \texttt{Local Region or Area} (8\%), \texttt{Celebrity} (7\%), \texttt{Health} (7\%), \texttt{Sports} (7\%), and \texttt{Finance} (5\%). 34\% wrote in a topic, with common themes including \texttt{Art/AI Art}, \texttt{Anime}, \texttt{Books}, \texttt{Education}, \texttt{Food}, \texttt{Memes}, \texttt{Pets}, and \texttt{Travel}.

\subsection{Behavioral Data}
To assess how founders' attitudes about their communities relate to the early trajectories of these spaces, we additionally computed a few aggregate community-level measures using behavioral data. Specifically, for each community referenced in a completed survey response, on the 28th day after the community's creation, we captured the cumulative number of unique visitors and contributors (post or comment), as well as the number of users subscribed to the community. Because community-building plans encompass both on- and off-platform actions, we did not collect data on specific actions taken within the communities to verify whether founders engaged in community-building plans.

\section{Founder Motivations for Community Creation}

\begin{table}[t]
\footnotesize
\centering
\begin{tabularx}{\textwidth}{Xcccc}
\toprule
\textbf{Scale / Item} &  &  &  & \\ 
\midrule
\rowcolor[HTML]{EFEFEF} 
\multicolumn{5}{l}{\textbf{Topical Interest} ($\alpha = 0.684$)} \\
\textit{I started this subreddit because the topic is entertaining.} & 0.89 & & & \\
\textit{I started this subreddit because the topic is exciting.} & 0.65 & & & \\
\midrule
\rowcolor[HTML]{EFEFEF} 
\multicolumn{5}{l}{\textbf{Connect with Others} ($\alpha = 0.758$)} \\
\textit{I started this subreddit to meet interesting people.} & & 0.65 & & \\
\textit{I started this subreddit to help me keep in touch with others.} & & 0.52 & & \\
\textit{I started this subreddit to enable people to express/share about themselves.} & & 0.47 & & \\
\textit{I started this subreddit to feel like I belong to a community.} & & 0.46 & & \\
\midrule
\rowcolor[HTML]{EFEFEF} 
\multicolumn{5}{l}{\textbf{Exchanging Information} ($\alpha = 0.706$)} \\
\textit{I started this subreddit to collect/provide information that might be useful.} & & & 0.81 & \\
\textit{I started this subreddit to share practical knowledge or skills.} & & & 0.62 & \\
\textit{I started this subreddit to enable discovery of new things.} & & & 0.53 & \\
\midrule
\rowcolor[HTML]{EFEFEF} 
\multicolumn{5}{l}{\textbf{Self-Promotion} ($\alpha = 0.811$)} \\
\textit{I started this subreddit because it helps me gain support or respect.} & & & & 0.84 \\
\textit{I started this subreddit because it helps me feel important.} & & & & 0.71 \\
\textit{I started this subreddit to build visibility in my personal or professional life.} & & & & 0.69 \\
\textit{I started this subreddit to get peer support from others.} & & & & 0.60 \\
\bottomrule
\end{tabularx}
\caption{Motivations for community creation on Reddit with constituent items and factor loadings.}
\vspace{-3em}
\label{tab:motivations}
\Description{This table summarizes the four factors describing our dimensions of motivation---topical interest, connect with others, exchanging information, and self-promotion. Each factor is presented with Cronbach alpha scores describing inter-item agreement, constitutent items, and the factor loadings of these items.}
\end{table}

To identify common motivations for founding communities on Reddit, we conducted exploratory factor analysis on 14 statements about founder motivations\footnote{Following Baglin~\cite{baglin2014improving}, we opt for polychoric over Pearson correlations for these ordinal measures.}. As our goal is to understand the motivations of founders intent on starting communities, we excluded from the factor analysis 99 respondents who strongly disagreed that ``It's important to me personally that this subreddit is successful''. Using the remaining 852 responses, we removed poorly-fitting items and refit factors until a satisfactory solution was achieved, aligning with prior CHI/CSCW methodologies~\cite{bentvelzen2021development, kairam2022social, mejia2017nine}. This resulted in the removal of a single poorly-fitting item `I started this subreddit because the topic is funny.'; the 13 remaining statements fit neatly into four distinct factors\footnote{The Kaiser-Meyer-Olkin test ($MSA = 0.88$) and Bartlett's test of sphericity ($p < 0.001$) confirmed the data's suitability for factor analysis. This solution yields four factors with eigenvalues exceeding 1, together accounting for 58\% of the variance in responses to items. All items have a loading score above 0.4.}, detailed in Table~\ref{tab:motivations}.

\begin{figure}[]
    \centering
    \includegraphics[width=\textwidth]{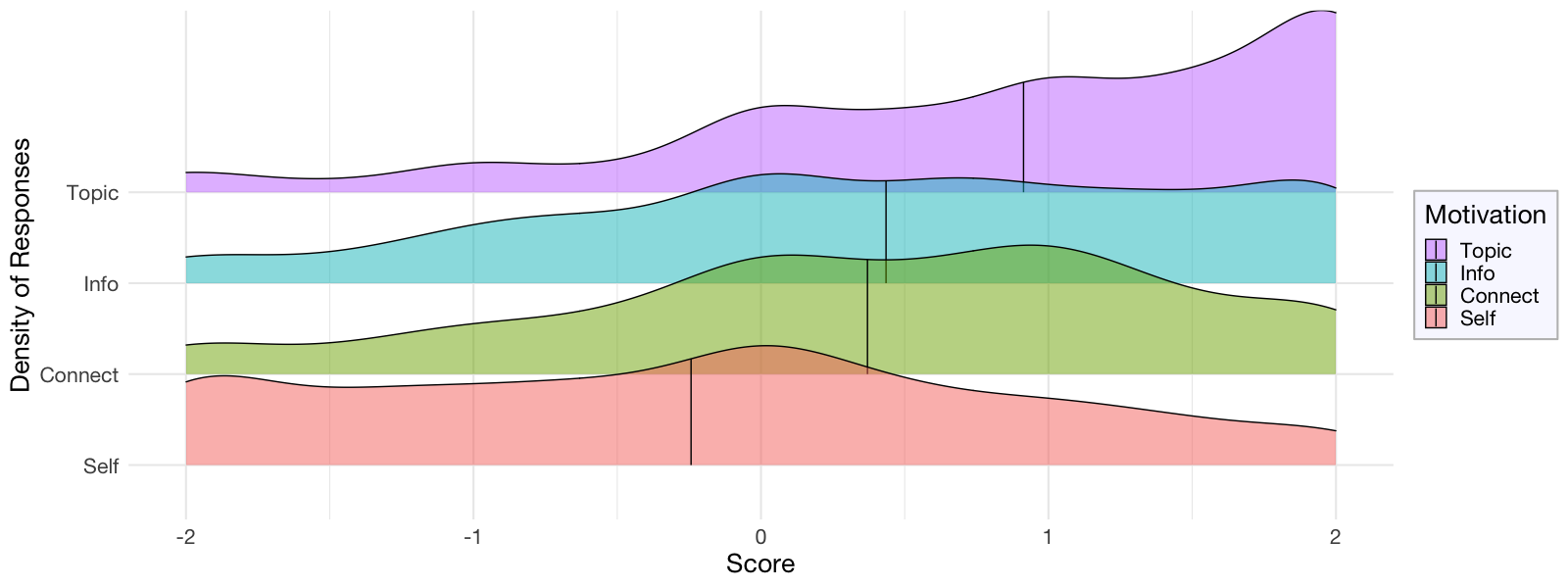}
    \vspace{-2em}
    \caption{Distribution among survey respondents of scores computed for each of the four motivations, on a scale from -2 (Strongly Disagree) to +2 (Strongly Agree). Lines show the mean for each motivation. Community creators were most likely to agree with topic engagement motivations ($\mu = 0.91$, $\sigma = 1.07$), followed by a motivation to exchange information ($\mu = 0.43$, $\sigma = 1.12$) or connect with others ($\mu = 0.37$, $\sigma = 1.06$). Self-promotion motivations were the least commonly reported ($\mu = -0.24$, $\sigma = 1.16$).}
    \label{fig:motivations-density}
    \Description{This figure shows a density ridge plot, summarizing the distributions of motivation scores across the full scale from -2 to +2. The distribution for `Topical Interest', shown at top in purple, is notably left-skewed, with a distribution centered at 0.91 and a peak close to the upper bound of +2. The distribution for `Self-Promotion', shown second in blue, has a peak at -0.24, and more probability mass in the negative (disagree) part of the scale. The third distribution, in green, is for `Connect with Others', with a peak at 0.91 and a plateau between 0 and +2. The distribution at the bottom, in red, shows `Exchange Info', centered roughly around a peak at 0.43.}
\end{figure}

Figure~\ref{fig:motivations-density} summarizes motivation score distributions for all respondents ($N = 951$). The top motivation for founding new communities on Reddit is \textbf{topical interest} ($\mu = 0.91$, $\sigma = 1.07$), capturing the extent to which founders are motivated by an `exciting' or `engaging' topic, with 63\% of surveyed founders agreeing or strongly agreeing with a typical item on this scale ($\mu >= 1.0$). The next most common motivations were \textbf{exchange information} ($\mu = 0.43$, $\sigma = 1.12$) or \textbf{connect with others} ($\mu = 0.37$, $\sigma = 1.06$), with 39\% and 37\% of creators agreeing / strongly agreeing with items on these scales, respectively. \textbf{Self-promotion} was the least frequently-reported motivation ($\mu = -0.24$, $\sigma = 1.16$), with just 19\% of creators agreeing / strongly agreeing with these items.

\section{Founder Criteria for Community Success}
Next, we explore success criteria for newly-created communities, based on respondents' relative rankings of seven potential measures of success, which are summarized in Table~\ref{tab:success-criteria}. Although each of the potential measures was ranked first by a substantial fraction of respondents, respondents broadly tended to favor quality-oriented over quantity-oriented measures. The measure ``A high level of interaction among community members'' had the highest average rank ($\mu$ = 3.27 of 7), while ``High quality information about the topic'' was most frequently ranked in the top slot (ranked \#1 by 25\% of respondents). Growth-oriented measures---a `large number of users' ($\mu$ = 3.64), `large volume of information' ($\mu$ = 3.82), `large number of contributors' ($\mu$ = 3.93)---tended to be ranked lower overall, but each was still ranked as the highest priority by a substantial fraction of respondents, indicating that creators have heterogeneous preferences regarding their communities. 129 respondents (13\%) chose to write in an option as ``Other'', with common themes including ``having fun'' and ``maintaining safety/respect''.

When asked to estimate what `fraction of people would potentially be interested in the content' in their new community, most creators did not expect their communities to have mass appeal. Only 9\% indicated that their community was for `Everyone', and just 23\% indicated that their community would be interesting to 50\% of people or more. 28\% indicated that their community was `very niche' and suitable for `only a handful of people', and 18\% chose `1-10\% of people'. The remaining 30\% of creators believed their community would have a potential audience of 10-50\% of people. 

\textbf{Success Criteria \& Motivations.} We analyzed relationships between founders' community creation motivations and how they envision success, using logistic regression to predict whether a founder's top-ranked motivation was Quality-Oriented (``high level of interaction'', ``high-quality information'', ``remains active for a long time'') or Quantity-Oriented (``large number of users'', ``large volume of information'', ``large number of contributors''), given scores for each of the four motivations. These are summarized in Table~\ref{tab:motivations-goals-reg}\footnote{Full distributions for how these measures were ranked are provided in the Appendix.}. Founders prioritizing quality-oriented success measures for their community had less `self-promotion' motivation and more `exchanging information' motivation. A one-point increase in responses to `self-promotion' items (e.g. moving from `Neutral' to `Agree') maps to a 27\% decrease in the odds of choosing a quality-oriented measure as the top criterion for success. Conversely, a one-point increase in `exchange information' responses maps to a 28\% increase in the odds of prioritizing a quality-oriented measure of success. 

\section{Founder Plans for Community Building}
Survey participants were asked if they planned to pursue several categories of actions to grow their communities, from raising awareness to rule enforcement; about half agreed that they would pursue each category. 45\% agreed or strongly agreed that they were ``pursuing specific actions or strategies to raise awareness about my subreddit.'' 53\% agreed / strongly agreed that they were ``pursuing actions or strategies to welcome new members'', 50\% ``to encourage contributions”, and 45\% ``to regulate behavior''. Notably, between 10-15\% of respondents ``strongly disagreed'' with each of these statements. 33\% of respondents indicated that they planned to devote 3 or more hours weekly to their community, while 42\% indicated that they planned to spend `1-3 hours per week'. The remaining 25\% of respondents planned to spend less than 1 hour per week.

\begin{table}
\footnotesize
\centering
\begin{tabularx}{\textwidth}{Xccc}
\toprule
\textbf{Success Measure} & \textbf{Avg. Rank} & \textbf{\% Top} & \textbf{\% Bottom 3} \\ 
\midrule
\rowcolor[HTML]{EFEFEF} 
\textit{A high level of interaction among community members} & 3.27 & 18\% & 25\% \\
\textit{High quality information about the topic} & 3.30 & 25\% & 32\% \\
\rowcolor[HTML]{EFEFEF} 
\textit{A community that remains active for a long time} & 3.55 & 15\% & 32\% \\
\textit{A large number of users viewing or consuming the content} & 3.64 & 16\% & 37\% \\
\rowcolor[HTML]{EFEFEF} 
\textit{A large volume of information about the topic} & 3.82 & 10\% & 40\% \\
\textit{A large number of contributors} & 3.93 & 11\% & 42\% \\
\rowcolor[HTML]{EFEFEF} 
\textit{Other (please explain)} & 6.48 & 5\% & 92\% \\
\bottomrule
\end{tabularx}
\caption{Success criteria for communities as ranked by founders (1 = Top-Ranked). Each option was ranked at the top by at least 10\% of founders (\% Top) and in the bottom 3 places by at least 25\% (\% Bottom 3).}
\vspace{-2em}
\label{tab:success-criteria}
\Description{This table summarizes how respondents ranked seven potential measures of success for their communities, on a scale from 1 (Top) to 7 (Bottom).}
\end{table}

\begin{table}[]
\centering
\begin{tabular}{@{}lS[table-format=-1.3]S[table-format=1.3]cS[table-format=1.3]@{}}
\toprule
& {$\beta$} & {SE} & {Odds Ratio [95\% CI]} & {$p$} \\ 
\midrule
Intercept & 0.200 & 0.096 & {---} & {---} \\
\midrule
Self-Promotion & -0.309 & 0.078 & 0.73 [0.63-0.86] & *** \\
Exchange Info & 0.250 & 0.074 & 1.28 [1.11-1.49] & *** \\
Connect with Others & 0.190 & 0.092 & [1.01-1.45] & {---} \\
Topical Interest & -0.136 & 0.068 & [0.76-1.00] & {---} \\ 
\bottomrule
\end{tabular}
\caption{Results of logistic regression model predicting whether founders would prioritize quality-oriented over quantity-oriented measures, based on expressed motivations. Coefficients ($\beta$) provided with standard errors (SE) and $p$ values (*: $p<0.01$, **: $p<0.005$, ***: $p<0.001$); we report odds ratios (OR), interpreted as the change in odds corresponding to a one-unit change in the variables, along with point estimates for predictors with $p < 0.01$. For example, a one-unit increase in self-promotion motivation (e.g. `Agree' vs. `Neutral') predicts a 27\% decrease in the odds of prioritizing quality-oriented measures of success.}
\label{tab:motivations-goals-reg}
\vspace{-2em}
\Description{This table summarizes results of a logistic regression model predicting founders' goals based on expressed motivations. The table has rows corresponding to the model intercept and four IVs, and columns capturing the beta coefficients, standard errors, computed odds ratios, and p-values.}
\end{table}

\textbf{Founder plans and motivations.} Differences in founders' likelihood of pursuing various actions can be predicted from self-reported motivations, as modeled using logistic regression. Results are summarized in Table~\ref{tab:motivations-plans-reg}. Stronger `self-promotion' and `meet others'  motivations map to increased odds of pursuing actions to raise awareness ($p < 0.001$). Stronger levels of all four motivations predict increased odds of pursuing actions to welcome newcomers ($p < 0.005$) and encourage contributions ($p < 0.01$). Stronger `exchange information' and `meet others' motivations predict increased odds of pursuing actions to regulate behavior in the community ($p < 0.005$). 

\begin{table}[]
\centering
\small 
\setlength{\tabcolsep}{4pt} 
\begin{tabular}{@{}l*{3}{S[table-format=-1.2]S[table-format=1.2]l}*{3}{S[table-format=1.2]S[table-format=1.2]l}@{}}
\toprule
 & \multicolumn{3}{c}{Raise Awareness} & \multicolumn{3}{c}{Welcome Newcomers} & \multicolumn{3}{c}{Encourage Contributions} & \multicolumn{3}{c}{Regulate Behavior} \\
\cmidrule(lr){2-4} \cmidrule(lr){5-7} \cmidrule(lr){8-10} \cmidrule(lr){11-13}
 & {$\beta$} & {SE} & {$p$} & {$\beta$} & {SE} & {$p$} & {$\beta$} & {SE} & {$p$} & {$\beta$} & {SE} & {$p$} \\
\midrule
Intercept & -0.56 & 0.11 & *** & -0.33 & 0.11 & ** & -0.50 & 0.11 & *** & -0.56 & 0.10 & *** \\
Self & 0.36 & 0.08 & *** & 0.22 & 0.08 & ** & 0.21 & 0.08 & * & 0.06 & 0.08 & {---} \\
Info & 0.15 & 0.08 & {---} & 0.26 & 0.08 & *** & 0.20 & 0.08 & * & 0.22 & 0.07 & ** \\
Connect & 0.48 & 0.10 & *** & 0.36 & 0.09 & *** & 0.38 & 0.10 & *** & 0.32 & 0.09 & *** \\
Topic & 0.19 & 0.08 & {---} & 0.26 & 0.07 & *** & 0.32 & 0.07 & *** & 0.16 & 0.07 & {---} \\
\bottomrule
\end{tabular}
\caption{Results of logistic regression models predicting whether founders intended to pursue actions within four categories, based on expressed motivations. Coefficients ($\beta$) provided with standard errors (SE) and $p$ values (*: $p<0.01$, **: $p<0.005$, ***: $p<0.001$)}
\label{tab:motivations-plans-reg}
\vspace{-3em}
\Description{This table summarizes results of four logistic regression models predicting whether founders' planned to take certain categories of actions in the community based on expressed motivations. The table has rows corresponding to the model intercept and four IVs, and columns capturing the beta coefficients, standard errors, and p-values. for each of the four categories of actions.}
\end{table}

\section{Founder Attitudes and Community Outcomes}
In this final section, we explore how differences in founders' motivations, goals, and plans predict early community trajectories, over the first 28 days. Regardless of whether founders prioritize quality over quantity long-term, communities must attract a critical mass of early participants to remain active~\cite{grudin1994groupware}. We thus evaluate early trajectories using three quantitative measures---total visitors, total contributors, and subscribers after 28 days. For each count-based outcome, we use negative binomial regression to capture how changes in the expected counts can be predicted based on differences in the independent variables of interest. For these models, to aid model convergence, we remove a single community that attracted over 150,000 unique visitors in the first 28 days.

\subsection{Founders' Motivations and Community Outcomes}
Differences in founder motivations predict substantial differences in early community trajectories, as summarized in Table~\ref{tab:motives-outcomes}. In particular, stronger `topical interest' motivations are associated with communities that attract consumers, contributors, and subscribers more quickly. Specifically, a one-point increase in average responses to items on the `topical interest' scale (such as moving from `Disagree' to `Neutral') is associated with a 124\% increase in the rate of attracting new visitors over the first 28 days ($\beta = 0.80$, \textit{Rate Ratio = 2.24}, $p < 0.001$), a 46\% increase in the rate of attracting new contributors ($\beta = 0.38$, $RR = 1.46$, $p < 0.001$), and a 131\% increase in the rate of attracting new subscribers ($\beta = 0.84$, $RR = 2.31$, $p < 0.001$).

\begin{table}
\centering
\small 
\setlength{\tabcolsep}{6pt} 
\begin{tabular}{@{}l*{3}{S[table-format=1.2]S[table-format=1.2]C[table-format=1.2]l}@{}}
\toprule
 & \multicolumn{4}{c}{Visitors} & \multicolumn{4}{c}{Contributors} & \multicolumn{4}{c}{Subscribers} \\
\cmidrule(lr){2-5} \cmidrule(lr){6-9} \cmidrule(lr){10-13}
 & {$\beta$} & {SE} & {RR} & {$p$} & {$\beta$} & {SE} & {RR} & {$p$} & {$\beta$} & {SE} & {RR} & {$p$} \\
\midrule
Intercept & 5.96 & 0.10 & {---} & *** & 1.17 & 0.07 & {---} & *** & 2.86 & 0.09 & {---} & *** \\
\midrule
Self-Promotion & 0.28 & 0.08 & 1.32 & *** & -0.16 & 0.05 & 0.85 & ** & -0.27 & 0.07 & 0.77 & *** \\
Exchange Info & -0.40 & 0.08 & 0.67 & *** & 0.05 & 0.05 & {---} & --- & -0.15 & 0.07 & {---} & --- \\
Connect with Others & -0.39 & 0.10 & 0.68 & *** & 0.15 & 0.06 & {---} & --- & -0.11 & 0.09 & {---} & --- \\
Topical Interest & 0.80 & 0.07 & 2.24 & *** & 0.38 & 0.05 & 1.46 & *** & 0.84 & 0.07 & 2.31 & *** \\
\bottomrule
\end{tabular}
\caption{Results from negative binomial regression models predicting differences in 28-day community outcomes associated with changes in founder motivations. Coefficients ($\beta$) are provided with standard errors (SE) and $p$ values (*: $p<0.01$, **: $p<0.005$, ***:$p<0.001$); for those with $p < 0.01$, we report exponentiated rate ratios (RR), interpreted as the change corresponding to a one-unit change in the variable in the rate at which the community gains users of the specified type. For example, a one-unit increase in the `exchange info' motivation predicts a 33\% decrease in the number of new visitors in the first 28 days.}
\label{tab:motives-outcomes}
\vspace{-2em}
\Description{This table summarizes results of four negative binomial regression models predicting 28-day values for the number of visitors, unique contributors, and total subscribers in a community, based on founders' expressed motivations.. The table has rows corresponding to the model intercept and four IVs, and columns capturing the beta coefficients, standard errors, computed rate ratios, and p-values. for each of the three outcome metrics of interest.}
\end{table}

Interestingly, stronger `self-promotion' motivations correlate with communities that attract more early viewers, but fewer contributors and subscribers. A one-point increase in average responses to `self-promotion' items is associated with a 32\% increase in new visitors to the community ($\beta = 0.28$, $RR = 1.32$, $p < 0.001$), a 15\% decrease in contributors ($\beta = -0.16$, $RR = 0.85$, $p < 0.005$), and a 23\% decrease in subscribers ($\beta = -0.27$, $RR = 0.77$, $p < 0.005$). Finally, increased `exchange info' and `connect with other' motivations are associated with a decrease in the rate of bringing in new viewers, but have no relationship with the number of contributors and subscribers. Specifically, a one-point increase in the `exchange info' motivation corresponds to a 33\% decrease in the rate of attracting new visitors ($\beta = -0.40$, $RR = 0.67$, $p < 0.001$), and a similar increase in the `meet others' motivation maps to a 32\% decrease ($\beta = -0.39$, $RR = 0.68$, $p < 0.001$).

\subsection{Founders' Success Criteria and Community Outcomes}
Founders who prioritize quality-oriented measures over quantity-oriented measures attract fewer unique visitors, but more contributors, over the first 28 days. Specifically, communities whose founders' top-ranked measure of success was quality-oriented attracted 59\% fewer visitors over the first 28 days ($\beta = -0.89$, $RR = 0.41$, $p < 0.001$), but 75\% more contributors ($\beta = 0.56$, $RR = 1.75$, $p < 0.001$), over the same period. A preference for quality-oriented measures did not predict any significant difference in the number of early subscribers.

\subsection{Community-Building Plans and Community Outcomes}
Differences in founders' plans for whether they would raise awareness, welcome newcomers, encourage contributions, and regulate behavior in their newly-created communities also predict differences in the early trajectories for their communities. In these models, as each plan was captured using a single 5-point Likert response, we represented these plans using binary variables capturing whether respondents `agreed' or `strongly agreed' that they intended to pursue actions or strategies within each of these categories. Our findings are summarized in Table~\ref{tab:actions-outcomes}.

Communities with founders who intended to pursue actions or strategies to raise awareness attracted more visitors, contributors, and subscribers within the first 28 days. Specifically, those who `agreed' or `strongly agreed' that they were pursuing such actions attracted 273\% more visitors ($\beta=1.32$, $RR = 3.73$, $p < 0.001$), 75\% more contributors ($\beta=0.56$, $RR = 1.75$, $p < 0.001$), and 189\% more subscribers ($\beta=1.06$, $RR = 2.89$, $p < 0.001$). Founders pursuing specific actions or strategies to welcome newcomers gained 58\% more unique contributors ($\beta=0.46$, $RR = 1.58$, $p < 0.001$), but did not attract significantly more visitors or subscribers, compared with founders who did not pursue these actions. Compared to those planning to raise awareness or welcome newcomers, founders with plans to encourage contributions interestingly saw a smaller increase in unique contributors ($\beta=0.34$, $RR = 1.40$, $p < 0.01$) and substantially fewer total visitors ($\beta=-0.82$, $RR = 0.44$, $p < 0.001$). Whether or not founders planned to regulate behavior had no discernible relationship with community activity within the first 28 days.

\begin{table}[]
\centering
\small 
\setlength{\tabcolsep}{4pt} 
\begin{tabular}{@{}l*{3}{S[table-format=1.2]S[table-format=1.2]C[table-format=1.2]l}@{}}
\toprule
 & \multicolumn{4}{c}{Visitors} & \multicolumn{4}{c}{Contributors} & \multicolumn{4}{c}{Subscribers} \\
\cmidrule(lr){2-5} \cmidrule(lr){6-9} \cmidrule(lr){10-13}
 & {$\beta$} & {SE} & {RR} & {$p$} & {$\beta$} & {SE} & {RR} & {$p$} & {$\beta$} & {SE} & {RR} & {$p$} \\
\midrule
Intercept & 5.85 & 0.11 & {---} & *** & 0.91 & 0.08 & {---} & *** & 3.25 & 0.10 & {---} & *** \\
\midrule
Raise Awareness & 1.32 & 0.19 & 3.73 & *** & 0.56 & 0.12 & 1.75 & *** & 1.06 & 0.17 & 2.89 & *** \\
Welcome Newcomers & 0.30 & 0.19 & {---} & {---} & 0.46 & 0.12 & 1.58 & *** & 0.42 & 0.17 & {---} & {---} \\
Encourage Contributions & -0.82 & 0.20 & 0.44 & *** & 0.34 & 0.13 & 1.40 & * & -0.24 & 0.18 & {---} & {---} \\
Regulate Behavior & 0.26 & 0.26 & {---} & {---} & 0.02 & 0.11 & {---} & {---} & -0.29 & 0.15 & {---} & {---} \\
\bottomrule
\end{tabular}
\caption{Results from negative binomial regression models predicting differences in 28-day community outcomes associated with founders' intentions (True/False) to pursue specific actions or plans within four categories: raising awareness, welcoming newcomers, encouraging contributions, and regulating behavior. Coefficients ($\beta$) are provided with standard errors (SE) and $p$ values (*: $p<0.01$, **: $p<0.005$, ***:$p<0.001$); for those with $p < 0.01$, we report exponentiated rate ratios (RR), interpreted as the change corresponding to a one-unit change in the variable in the rate at which a community gains users of the specified type. For example, communities with founders who report specific plans to raise awareness attract 273\% more visitors over the first 28 days than founders who don't.}
\label{tab:actions-outcomes}
\vspace{-2em}
\Description{This table summarizes results of four negative binomial regression models predicting 28-day values for the number of visitors, unique contributors, and total subscribers in a community, based on founders' planned actions. The table has rows corresponding to the model intercept and four IVs, and columns capturing the beta coefficients, standard errors, computed rate ratios, and p-values. for each of the three outcome metrics of interest.}
\end{table}

\section{Discussion}
This paper provides the first detailed account of the attitudes with which online community founders approach their new communities, drawing on data from a survey of 951 recent founders of a diverse set of communities across Reddit. We find that founder motivations are grouped into four dimensions---topical interest, exchanging information, connecting with others, and self-promotion. These dimensions align almost exactly with the prior studies of contributor motivations from which survey items were drawn~\cite{moore2017redditors, park2009being} and echo findings that `spreading information and building community' around a topic represent primary motivations for wiki founders~\cite{foote2017starting}.

Among the founders we studied, `topical interest' motivations were most prevalent, and `self-promotion' least prevalent. This contrasts with prior studies identifying stronger `exchange information' and `connect with others' motivations for lurkers who become contributors~\cite{ridings2006psychological, lampe2010motivations, wang2003assessing}. One explanation is that `topical interest' means something different for founders than for participants in a community. For founders, `topical interest' relates to an active process of creating and curating a social space around a particular topic; for participants, `topical interest' may align more with passive consumption, and thus with `entertainment' motivations identified in prior work. 

Interestingly, founders motivated by `topical interest' create communities that attract more early visitors, contributors, and subscribers. These findings echo prior work which identifies how posters in microblogging systems who maintain a narrower topical focus attract an audience more quickly~\cite{wang2012twitter}. A possible explanation is that these communities, centered around a specific topic of interest, create a clearer value proposition for potential members who are learning about the community or who are determining if and how they should contribute. A focused topic that brings together people with common interests may further support the building of identity-based commitment to the community among members~\cite{kraut2012building,ren_building_2012, mcpherson2001birds}.

While founders' goals for their nascent communities varied, founders generally prioritized quality over quantity as a measure of success. This finding aligns with previous research showing that participants and moderators prioritize quality over size within established communties~\cite{weld_what_2021,hwang_why_2021, foote2017starting}. Founders motivated to `exchange information' were more likely to prioritize community quality, while those motivated by `self-promotion' were more likely to prioritize quantity. Ironically, stronger `self-promotion' motivations were associated with communities that gained fewer early contributors and subscribers. These may represent founders who view their subreddit more as an `audience' or `fan community' for their work (as in~\cite{shamma2009spinning}) and thus create a more one-sided expectation in which they are the contributors and community members are the consumers.

Finally, we explored four categories of actions associated with successful communities~\cite{kraut2012building}: raising awareness, welcoming newcomers, encouraging contributions, and regulating behavior, finding that founders expressed intention to pursue these categories of actions at roughly equal rates. Having plans to `raise awareness' was the largest determinant of a community's ability to attract early visitors, contributors, and subscribers. Plans to `welcome newcomers' had no relationship to the number of visitors, but did increase the number of contributors; in the context of Reddit, contributions capture all social interactions, and prior research has highlighted how efforts to welcome newcomers can encourage subsequent interactions~\cite{kraut2012building, burke2009feed, arguello2006talk, lampe2005follow}. In contrast, having specific plans to `encourage contributions' was actually associated with fewer visitors; one possible explanation is that additional pressure to contribute could create an `entry barrier', which dissuades some newcomers who are less interested or committed to the group~\cite{kraut2012building}.

\subsection{Opportunities for Designers and Founders}
Broadly, this work shows that founders of online communities approach their projects with varied motivations and goals, translating into different patterns of activities, which then shape the early trajectories of their communities. Founders of online communities on services outside of Reddit, such as Discord or Facebook Groups, are likely to vary in the motivations, topics, and actions associated with their communities; we have thus prioritized opportunities below that we believe are likely to generalize across platforms, as they focus more on adapting the design of these systems based on the motivations, topics, lifecycle stage, and goals associated with a community.

\textbf{Matching founder motivations.} Varied founder motivations create opportunities for services to move beyond a `one-size-fits-all' approach to community creation. For example, platforms could customize onboarding and tooling based on founders' expressed motivations. Founders motivated by `exchanging information', for instance, might benefit from collaborative workspaces (e.g., wikis) and summarization tools, while those motivated to `connect with others' may privilege conversational threads. Similarly, based on our observation that founders and contributors express motivations on similar dimensions, it may be possible to use information about stated or inferred founder motivations to help customize discovery surfaces, in order to help direct individuals to communities that better match their motivations. 

\textbf{Providing topic-specific support.} We find that founders motivated by a specific topic tended to have communities with more successful early trajectories. For founders without a clear topical focus, it may be helpful to `nudge' them towards a specific topical niche, to help increase their odds of success. Future research could also explore topically-relevant onboarding and support. There may be certain kinds of early community-building actions that are best suited for gaming communities, for instance, or other actions suited for communities focused on current events.

\textbf{Designing for community lifecycles.} Our research also identified that certain categories of actions were particularly valuable for nascent communities. Tools making it easier for founders to raise awareness about their communities, for instance, could potentially help more communities successfully `launch' by attracting early visitors, contributors, and subscribers. Conversely, based on our finding that early plans to encourage contributions may actually reduce the number of visitors to a community, we might design onboarding that encourages community creators to wait longer before ramping up calls to contribute. Further research could help to identify when in their lifecycle communities can most successfully transition to encouraging broader contribution, adding moderation, and implementing other community-building strategies.

\textbf{Visualizing community success.} Finally, our finding that founders varied broadly in their measures of success for their communities highlights an opportunity for providing more customized analytics to help founders track the progress of their communities. While many platforms make readily available quantity-oriented metrics, such as the number of subscribers or contributions, researchers and designers should consider opportunities for visualizing the `quality' of a community. How might we characterize the quality of information or conversational interaction for community founders and moderators in a way that is actionable? We envision a rich design space here for supporting the majority of community founders who are pursuing these goals.

\subsection{Limitations \& Opportunities for Future Work}
Before concluding, we consider some limitations of our findings and opportunities for addressing these in future research. Respondents self-selected into our survey, raising concerns about non-response bias. While our approach allows us to identify the range of motivations and goals associated with a diverse set of communities, it is not a reliable method of estimating the overall distribution of these attitudes. In particular, we almost certainly under-sampled users who created communities with very low levels of motivation (e.g., as a way of exploring the site \cite{foote_behavior_2018}). 

Future work employing larger, more representative samples could not only answer these questions, but also identify patterns of motivations associated with important dimensions of identity, such as race/ethnicity or sexual orientation, which are likely to influence founders' motivations and expectations for their communities. In addition, we note that this survey was conducted among only English-speaking community founders; an exciting area for future research is understanding whether and how these motivations and goals for founding online communities vary across cultures and languages, and how these differences impact early community growth and success.

The aim of this study was to understand how founders' attitudes relate to the initial success of their communities; future work could explore predictors of long-term community success. This study relied on founders' self-reported plans for community building activities; future work could leverage behavioral logs to analyze the extent to which specific on-platform actions within these categories (e.g. posting a welcome message, adding community rules) drive community success (as in~\cite{kraut2014role}). Content analysis, particularly of founders' own contributions, might yield further insight into questions raised in this paper, such as characterizing how the specific language founders use to encourage contributions might discourage certain community members.


Finally, we have focused in this research on Reddit as it provided access to founders with a broad range of topics and use cases for their communities. We find many similarities to findings from prior research on founders of wiki projects~\cite{foote2017starting}, indicating that these insights may generalize across other types of online community and co-production platforms. We encourage future research that explores more fully how founders' motivations and goals, and the impact of these attitudes on early community success, vary across platforms with different affordances, policies, or cultural norms.

\section{Conclusion}
In this short paper, we have presented results from a survey of founders of online communities soon after the creation of their communities, allowing us to characterize their motivations and goals before they have been shaped by their experiences moderating these spaces. We show that differences in these founder attitudes correspond to differences in the subsequent trajectories of these communities, illustrating the substantial role that founders play in shaping these spaces, especially early on. With this work, we hope to encourage research and design to support a more diverse set of founders in achieving their goals and creating valuable resources for others through the work of creating and leading online communities.

\begin{acks}
We wish to acknowledge and thank first the community founders who shared their experiences with us as part of this research. We wish to thank Zareen Ahmad Brock, Ashlee Edwards, Jason Fong, Carl Pearson, Michelle Quin, Jeff Robertson, Cruza Shamblin, and Bradley Spahn for their support in designing and deploying this study. We also thank our CHI reviewers, whose thoughtful feedback greatly improved this work.
\end{acks}


\bibliographystyle{ACM-Reference-Format}
\bibliography{sample-base}

\pagebreak
\appendix

\section{Survey Measures}
The following measures were included in the survey. Likert-scale responses were given using a 5-point scale from \textit{Strongly Disagree} to \textit{Strongly Agree}. Item order was randomized within questions.

\small
\begin{itemize}
    \item What is your age?
    \begin{itemize}
        \item Under 15 [Survey terminates]
        \item 16-17 [Survey terminates]
        \item 18-24
        \item 25-34
        \item 35-44
        \item 45-54
        \item 55-64
        \item 65+
    \end{itemize}
    \item How long have you been active on Reddit?
    \begin{itemize}
        \item I visit Reddit multiple times per day
        \item Once per day
        \item A few times per week
        \item A few times per month
        \item Once per month or less
    \end{itemize}
    \item Have you ever started or moderated an online community or collaborative project on any service outside of Reddit?
    \begin{itemize}
        \item Yes
        \item No
        \item I'm not sure
    \end{itemize}
    \item Have you created or moderated any other subreddit on Reddit, other than the one you recently created (r/<subreddit>)? Please choose the response that best applies to you:
    \begin{itemize}
        \item Yes, I have previously created a subreddit.
        \item Yes, I have previously moderated a subredit, but never created one.
        \item No, I have not moderated or created a subreddit before this one.
    \end{itemize}
    \item Which of the following represents the intended topic for r/<subreddit>? Please choose any answer(s) that would describe the content within your subreddit as you have imagined it. If no option matches, please choose "Other" and provide a short description of the intended topic.
    \begin{itemize}
        \item Entertainment
        \item Technology
        \item Music
        \item Humor
        \item Local Region or Area
        \item Mature Themes \& Adult Content
        \item Sports
        \item Celebrity
        \item Medical or Mental Health
        \item News/Politics
        \item Finance
        \item Other (please describe)
    \end{itemize}
    \item Which of the following permitted subreddit ratings would best characterize the intended content within r/<subreddit>? Please choose the answer that best applies.
    \begin{itemize}
        \item E [Everyone]: Occasionally or never posts and discusses mature themes.
        \item M [Mature]: Regularly posts and discusses mature themes.
        \item D [High-Risk Drug Use]: Regularly posts and discusses drug use with a high risk of being dangerous or harmful.
        \item V [Violence \& Gore]: Regularly posts and discusses graphic violence, gore, or surgical procedures.
        \item X [Sexually Explicit]: Regularly posts and discusses explicit sexual content or pornography.
    \end{itemize}
    \item Please think about your initial goals for r/{<subreddit>} and rate your level of agreement or disagreement with each of the following statements:
    \begin{itemize}
        \item I started this subreddit to get peer support from others.
        \item I started this subreddit to meet interesting people.
        \item I started this subreddit to feel like I belong to a community.
        \item I started this subreddit to help me keep in touch with others.
        \item I started this subreddit because the topic is entertaining.
        \item I started this subreddit because the topic is funny.
        \item I started this subreddit because the topic is exciting.
        \item I started this subreddit because the topic is enjoyable.
        \item I started this subreddit to collect/provide information that might be useful.
        \item I started this subreddit to share practical knowledge or skills.
        \item I started this subreddit to enable discovery of new things.
        \item I started this subreddit to enable people to express/share about themselves.
        \item I started this subreddit because it helps me gain support or respect.
        \item I started this subreddit to build visibility in my personal or professional life.
        \item I started this subreddit because it helps me feel important.
    \end{itemize}

    \item Think about how you will assess whether or not your community is successful, and rank which of the following are the most important measures of success for you. [ranked choice]:
    \begin{itemize}
        \item A large number of contributors
        \item A large volume of information about the topic
        \item High quality information about the topic
        \item A community that remains active for a long time
        \item A high level of interaction among community members
        \item A large number of users viewing or consuming the content
        \item Other (please explain)
    \end{itemize}

    \item How much time are you planning to dedicate to your subreddit?
    \begin{itemize}
        \item Less than an hour a week
        \item 1-3 hours per week
        \item 3-10 hours per week
        \item More than 10 hours per week
    \end{itemize}

    \item Please rate your level of agreement with each of the following statements:
    \begin{itemize}
        \item I am using specific actions or strategies to raise awareness about my subreddit.
        \item I am using specific actions or strategies to welcome new members to my subreddit.
        \item I am using specific actions or strategies to encourage new members to contribute to my subreddit.
        \item I am using specific actions or strategies to regulate the behavior of new members.
        \item It’s important to me personally that this subreddit is successful.
    \end{itemize}    

    \item How do you identify? If you prefer to self-describe, please do so in the box below marked "Other."
    \begin{itemize}
        \item Male
        \item Female
        \item Non-binary/third gender
        \item Prefer not to answer
        \item Other (please specify):
    \end{itemize}
\end{itemize}

\pagebreak
\section{Full Founder Ranking of Success Criteria}
\normalsize
Here, we provide the full distributions of ranks assigned by founders for each of the proposed measures of community success, with 1 indicating the highest rank and 7 indicating the lowest.

\begin{figure}[h]
    \centering
    \includegraphics[width=.8\linewidth]{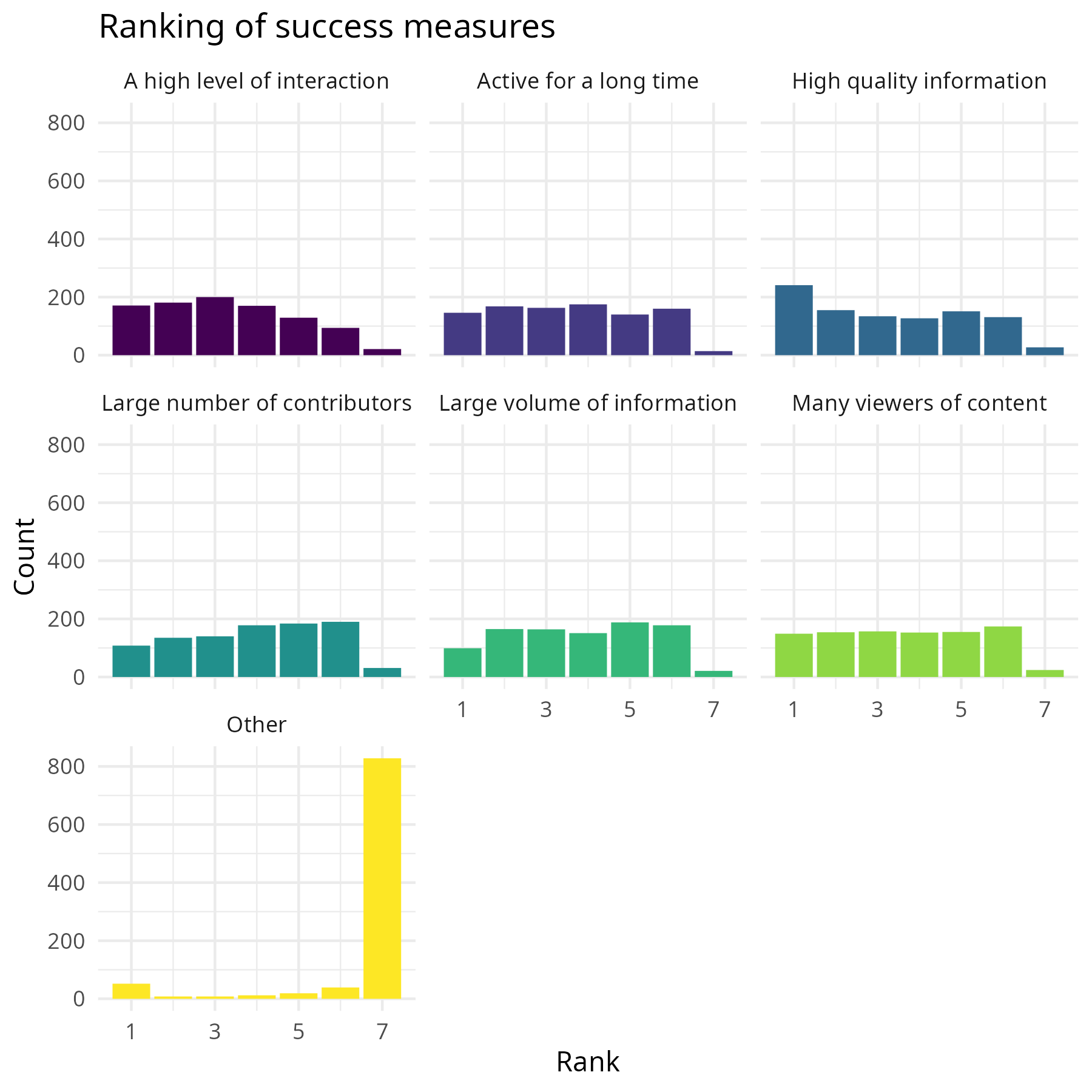}
    \caption{The ranking given to each of the measures of success provided. There is a large amount of heterogeneity in rankings, and each measure receives many high and low rankings.}
    \Description{This faceted chart shows the distribution of ranks for each of the 6 proposed criteria for community success, along with ranks for the `Other' category. For each of the success measures, the distribution across ranks 1 to 6 is close to uniform, while `Other' predominates in the lowest (7) rank.}
    \label{fig:success_ranking}
\end{figure}


\end{document}